\documentstyle[12pt]{article}

\begin{document}
\setlength{\textheight}{1.2\textheight}


\def\beqn{\begin{equation}}
\def\eeqn{\end{equation}}
\def\beqa{\begin{eqnarray}}
\def\eeqa{\end{eqnarray}}



\def\epaisfle{height1pt}
\def\longfle{width0.5cm}
\def\blancfle{\vskip2pt}
\def\espfle{\vskip0.2cm}
\def\espflc{\vskip0.3cm}

\def\flecher{\hbox{\vbox{\hrule \longfle \epaisfle
\blancfle}$\triangleright$}}

\def\flechel{\hbox{$\triangleleft$\vbox{\hrule \longfle \epaisfle
\blancfle}}}

\def\fleches{\vbox{\flecher \espfle \flechel \espflc}}


\def\epaisdir{height0.3pt}
\def\blancdir{\vskip2.4pt}


\def\largmir{width0.2cm}
\def\epaishmir{height1pt}
\def\epaisvmir{width1pt}
\def\hautmir{height1.3cm}
\def\intmir{\hskip0.13cm}

\def\miroir{\vbox{\hrule \largmir \epaishmir
\hbox{\vrule \hautmir \epaisvmir \intmir \vrule \hautmir \epaisvmir}
\hrule \largmir \epaishmir}}

\def\espace{\hskip0.15cm}


\def\phn{\varphi_{in\;\>}}
\def\pho{\varphi_{out}}
\def\psn{\psi_{in\;\>}}
\def\pso{\psi_{out}}
\def\phc{\varphi_{cav}}
\def\psc{\psi_{cav}}
\def\espvca{\vskip1.23cm}
\def\esphcav{\hskip0.3cm}
\def\longflc{width1cm}
\def\longcav{width1.3cm}
\def\espq{\hskip0.7cm}

\def\champl{\hbox{\vbox{\hbox{$\phn$} \espfle \hbox{$\pso$} \espflc}
\fleches}}
\def\champr{\hbox{\fleches \espace
\vbox{\hbox{$\pho$} \espfle \hbox{$\psn$} \espflc}}}

\def\flechcr{\hbox{\vbox{\hrule \longflc \epaisfle
\blancfle}$\triangleright$}}
\def\flechcl{\hbox{$\triangleleft$\vbox{\hrule \longflc \epaisfle
\blancfle}}}
\def\direcc{\hbox{$\langle$\vbox{\hrule \longcav \epaisdir
\blancdir}$\rangle$}}
\def\champc{\vbox{\hbox{\espace \vbox{\hbox{\esphcav $\phc$} \flechcr
\espfle \flechcl \hbox{\esphcav $\psc$}}} \direcc \hbox{\espq $q$}}}

\def\MIROIR{
\hbox{\champl \espace \miroir \espace \champr}
}

\def\CAVITE{
\hbox{\vbox{\hbox{\champl \espace \miroir} \espvca}\champc
\vbox{\hbox{\miroir \espace \champr} \espvca}}
}


\def\q{\dot q}
\def\dddq{\stackrel{...}{\delta q}}
\def\dq{\delta q}
\def\dvphi{\dot \varphi}
\def\dpsi{\dot \psi}

\begin{titlepage}
\nopagebreak
\begin{center}
{\large\bf MECHANICAL EFFECTS OF RADIATION PRESSURE QUANTUM FLUCTUATIONS}\\

 \vfill
        {\bf Marc-Thierry Jaekel\dag ~and Serge Reynaud\ddag }  \\
\end{center}
\dag Laboratoire de Physique Th\'eorique
 de l'Ecole Normale  Sup\'erieure\footnote{Unit\'e
propre du Centre National de la Recherche Scientifique, \\
associ\'ee \`a l'Ecole Normale Sup\'erieure et \`a l'Universit\'e
de Paris Sud.}(CNRS),
24 rue Lhomond, F75231 Paris Cedex 05, France \\
\ddag Laboratoire Kastler-Brossel\footnote{Unit\'e
de l'Ecole Normale Sup\'erieure et de l'Universit\'e Pierre et Marie Curie,\\
 associ\'ee au Centre National de la Recherche
Scientifique.}(UPMC-ENS-CNRS), case 74,\\
 4 place Jussieu,  F75252 Paris Cedex 05,
France \\

\vfill

\vfill

\begin{abstract}
As revealed by space-time probing, mechanics and field theory come out as
complementary descriptions for motions in space-time. In particular,
quantum fields
exert a radiation pressure on scatterers which results in
mechanical effects that persist in vacuum. They include mean forces due to
quantum field fluctuations, like Casimir forces, but also
fluctuations of these forces and additional forces linked to
motion. As in classical electron theory, a moving scatterer
is submitted to a radiation reaction
force which modifies its motional response to an applied force.
We briefly survey the mechanical effects of quantum field fluctuations
and discuss the consequences for stability of motion in vacuum and for position
fluctuations.
\bigskip

\end{abstract}
\vfill

\begin{flushleft}

\vfill
        {\normalsize LPTENS 95/5\\
May 1995\\}

\vfill

Proceedings of NATO-ASI Conference "Electron Theory and Quantum Electrodynamics
- 100
years later", September 5-16, 1994, Edirne, Turkey.
\end{flushleft}

\end{titlepage}

\begin{flushleft}
{\bf 1 Introduction}
\end{flushleft}

Lorentz electron theory \cite{Lorentz} was an early unification of fields and
particles, in that case electromagnetic fields and charged particles,
in a common and universal description. This frame played a determinant role
in a consistent
development of classical field theory and relativistic mechanics \cite{E1}.
This close connection was deeply perturbed by the advent of quantum formalisms,
which ultimately emphasize the primary role of quantum fields.
Within the framework of quantum electrodynamics, mechanical
effects on charged particles, although obtainable in principle, are
usually derived with difficulties \cite{QED}.

When taking into account physical limits in space-time probing,
field theory and mechanics emerge as
complementary representations in space-time.
Fields are measured by their mechanical effects on test particles,
while particle positions are measured through probe fields.
Space-time measurements are determined by energy-momentum exchanges
between fields and particles. In the same way as radiation of test
particles affects field measurements, mechanical effects of the probe field
radiation pressure affect position measurements, and hence
the determination of motions in space-time.

Quantum fluctuations impose ultimate limitations which affect both
particle positions and field values in space-time \cite{BR}.
As a consequence of Heisenberg inequalities, oscillators have
fluctuations which subsist in their ground states.
Quantum field fluctuations persist in vacuum, and those vacuum fluctuations
result in fundamental limitations on the determination not only of fields,
but also of positions and motions in space-time \cite{Braginsky}.
The presence of ultimate
quantum fluctuations must then be taken into account
in a consistent development of
mechanics.

Macroscopic objects feel the effects of quantum field fluctuations,
even in vacuum.
A well-known example is that of Casimir forces between macroscopic conductors
or dielectrics \cite{Casimir}. Casimir forces can be seen as a
mechanical signature of the radiation pressure of quantum field fluctuations.
But radiation pressure itself presents quantum fluctuations, which result
in further mechanical effects. When moving in a fluctuating environment,
a scatterer radiates and experiences a radiation reaction force
\cite{E2}.
There results a typical quantum Brownian motion which is determined
by radiation pressure fluctuations and which persists in vacuum.
We shall here briefly survey these effects,
and also discuss the fundamental consequences for stability of
motion in vacuum and for ultimate fluctuations of position.

\bigskip
\begin{flushleft}
{\bf 2 Radiation pressure of quantum field fluctuations}
\end{flushleft}

Reflectors which are immersed in a fluctuating field,
like conductors or dielectrics
in a fluctuating electromagnetic field for instance, feel the effects of
field fluctuations. Even at the limit of zero temperature, i.e.
in a field which does not contain any photon,
that is in the vacuum state, effects of quantum field fluctuations persist.
Vacuum fluctuations are responsible for a mean force, the so-called Casimir
force, between reflectors. Two plane infinite perfect mirrors
in the electromagnetic vacuum are submitted to a mean attractive pressure
 (force per unit area), which
decreases like the fourth power of the mirrors' distance $q$ ($\hbar$ is
Planck constant, $A$ the area, light velocity is equal to $1$):
\beqn
\label{fc}
F_c = { \pi^2 \over 240}{\hbar  \over q^4} A
\eeqn
Although very weak, Casimir forces between macroscopic plates have
effectively been observed \cite{Casimir}. As a useful simple illustration,
we shall also discuss in the following
the similar system made of two perfect
mirrors in two-dimensional space-time, in the vacuum of a scalar field.
In that case, the mean force decreases like the second power of the mirrors'
distance:
\beqn
\label{fc2}
F_c = {\pi  \over 24}{ \hbar  \over q^2}
\eeqn
For perfect mirrors, the mean Casimir force can easily be derived,
using the following
simple argument \cite{Casimir}.
 The field can be considered as a set of free oscillators
in each of the space domains delimited by the mirrors.
Each of these oscillators possesses (zero-point) quantum fluctuations
in its ground state, with a corresponding energy of
 ${1\over2} \hbar \omega$ ($\omega$ the oscillator's frequency).
The mirrors play
the role of boundary conditions for the field, modifying the spectrum of
mode frequencies allowed in the cavity they form. The total zero-point energy
of the field
($\sum_n {1\over2} \hbar \omega_n$) varies with the mirrors' distance,
and results in the mean force (\ref{fc}) between the mirrors. Although
it clearly exhibits the role of vacuum field fluctuations,
this simple interpretation in terms of vacuum energy leads to problems.
 Because of its high frequency
dependence, the mode spectrum results in a total zero-point energy which is
infinite, or which at least corresponds to a high energy
density that induces problematic gravitational and cosmological
consequences \cite{Wesson}. On another hand, when comparing different
physical situations, variations of vacuum energy (with the distance for
instance) produce finite and observable effects. Moreover, realistic mirrors
must clearly be transparent to sufficiently high field frequencies,
whose fluctuations should then have no incidence on the effect.

An alternate and more consistent derivation of Casimir forces uses
a local description in terms of the radiation pressure exerted by
field fluctuations on the mirrors \cite{BM}. We briefly recall this derivation
in the case of the simple model of mirrors in two-dimensional
space-time ($(x^\mu)_{\mu=0,1} = (t,x)$) (\ref{fc2}).
Each mirror determines two regions of space where
the scalar field $\phi$ can be decomposed
on two components which propagate freely
(Figure 1):
$$\phi(t,x) = \varphi(t-x) + \psi(t+x)$$
A perfect mirror corresponds to a boundary condition for the field,
saying that the field vanishes at the
mirror's position in space ($q$):
\beqn
\label{bc}
\varphi_{out}(t-q) = - \psi_{in}(t+q) \qquad \qquad
\psi_{out}(t+q) = - \varphi_{in}(t-q)
\eeqn
{}From now on, we shall use the following notation for Fourier transforms:
$$f(t) =  \int_{-\infty}^\infty {d\omega \over 2\pi}  e^{-i\omega t}
f[\omega]$$
In a more realistic model, the mirror partially reflects and transmits
both components of the field and is described by a scattering matrix
\cite{JR1}:

\bigskip
\hskip4.3cm \MIROIR
$$Figure \quad 1$$
\beqn
\label{S}
\left(\matrix{\varphi_{out}[\omega]\cr
\psi_{out}[\omega]\cr}\right) = S[\omega]
\left(\matrix{\varphi_{in}[\omega]\cr
\psi_{in}[\omega]\cr}\right) \qquad \qquad
S[\omega] = \left(\matrix{s[\omega]&r[\omega]e^{-2i\omega q}\cr
r[\omega]e^{2i\omega q}&s[\omega]\cr}\right)\nonumber\\
\eeqn
The mirror is assumed to be very heavy when compared with the field energy,
so that under reflection momentum is transfered to the field while its energy
is preserved (the mirror's recoil is neglected).
$s$ and $r$ are frequency dependent transmission and reflection
amplitudes, and must obey the general analyticity and unitarity conditions
of scattering matrices which correspond to causality of field scattering and
conservation of probabilities. In addition, we shall assume high frequency
transparency, i.e. that reflection coefficients vanish sufficiently
rapidly when frequency goes to infinity.

Two mirrors form a
Fabry-Perot cavity which divides space into three domains where the field
propagates freely (Figure 2).

\bigskip
\hskip3.6cm \CAVITE
$$Figure \quad 2$$
All fields are determined by input fields and
the mirrors' scattering matrices. On each side of each mirror, the radiation
pressure exerted by the field is provided by the field stress tensor
($T^{\mu\nu}$):
\beqa
\label{st}
T^{00} &=& T^{11} = {1\over2}(\partial_t \phi^2 +
\partial_x \phi^2)\nonumber\\
T^{01} &=& T^{10} = - \partial_t \phi
\partial_x \phi
\eeqa
The force is given by the flux of stress tensor component $T^{11}$ through the
mirror (difference between left and right sides; a dot stands for time
derivative):
\beqa
\label{f}
F_1 &=&  \lbrace \dvphi_{in}(t-q_1)^2 +
\dpsi_{out}(t+q_1)^2\rbrace - \lbrace\dvphi_{cav}(t-q_1)^2 +
 \dpsi_{cav}(t+q_1)^2
\rbrace\nonumber\\
F_2 &=&  \lbrace \dvphi_{cav}(t-q_2)^2 +
\dpsi_{cav}(t+q_2)^2\rbrace - \lbrace\dvphi_{out}(t-q_2)^2 +
 \dpsi_{in}(t+q_2)^2
\rbrace\nonumber\\
\eeqa
$<F_1> = - <F_2> = F_c$ is the mean radiation pressure,
or Casimir force, felt by the mirrors \cite{JR1}.

In vacuum state, the two input field components are uncorrelated
and have identical auto-correlations:
\beqn
\label{vac}
<\varphi_{in}[\omega]\varphi_{in}[\omega']> =
<\psi_{in}[\omega]\psi_{in}[\omega']> = {2\pi \over\omega^2}
\delta(\omega+\omega')\theta(\omega) {1\over2} \hbar \omega
\eeqn
($\theta$ is Heaviside's step function).
Although the corresponding vacuum mean energy density is infinite
(the spectral energy density increases linearly with frequency),
the mean forces exerted on the mirrors (\ref{f}) are finite, as the
reflection coefficients of the mirrors ($r_1$ and $r_2$)
satisfy high frequency transparency ($q = q_2-q_1$):
$$F_c = \int_0^\infty {d\omega \over 2\pi} \hbar \omega
\lbrace 1 - g[\omega]\rbrace$$
$$g[\omega] = {1 - |r[\omega]|^2 \over |1 - r[\omega]e^{2i\omega q}|^2}
\qquad \qquad r[\omega] = r_1[\omega] r_2[\omega]$$
$g$ describes the field spectral energy density inside the cavity.
At the limit of perfect mirrors ($r = 1$), $g$ becomes a sum of delta
functions at frequencies equal to the modes of the cavity ($n {\pi\over q}$)
and the Casimir force between the two mirrors identifies with (\ref{fc2}).

\bigskip
\begin{flushleft}
{\bf 3 Quantum fluctuations of radiation pressure}
\end{flushleft}

The radiation pressure exerted on a scatterer
is related to the field stress tensor,
so that it is a function of fields
(in general a quadratic form, see (\ref{st}) for instance).
Consequently, quantum fluctuations of fields also induce
quantum fluctuations of stress tensors and radiation pressures.
The fluctuations of Casimir forces exerted on mirrors,
due to quantum fluctuations
of electromagnetic fields, have recently been studied \cite{Barton}.
We shall just discuss some general properties of
stress tensor and radiation pressure fluctuations in vacuum.

Electromagnetic fields $F_{\mu\nu}$ and stress tensor components
$T_{\mu\nu}$ can be
derived from electromagnetic potentials $A_\mu$
($F_{\mu\nu} = \partial_\mu A_\nu - \partial_\nu A_\mu $,
$\eta_{\mu\nu}$ is
Minkowski metric $diag(1,-1,-1,-1)$):
$$T_{\mu\nu} = {F_\mu}^\lambda F_{\nu\lambda} - {1\over 4}
\eta_{\mu\nu} F^{\rho\lambda} F_{\rho\lambda}$$
Vacuum correlations of electromagnetic potentials are determined from
propagation equations (in Feynman gauge):
\beqa
<A_\mu(x) A_\nu(0)> &=&
\int{d^4k \over (2\pi)^4} e^{-ik.x} C_{A_\mu A_\nu}[k]\nonumber\\
C_{A_\mu A_\nu}[k] &=& 2\pi \hbar \theta(k_0) \delta(k^2)\eta_{\mu\nu}\nonumber
\eeqa
These expressions exhibit translation and Lorentz invariances, and a spectrum
limited to light-like momenta with positive frequencies
(as vacuum is the state of minimum energy,
transitions only have positive frequencies).
In vacuum, correlations of stress tensors are determined from
field correlations using Wick'rules:
$$<T_{\mu\nu}(x)T_{\rho\sigma}(0)> - <T_{\mu\nu}(x)><T_{\rho\sigma}(0)>
= \int{d^4k \over (2\pi)^4} e^{-ik.x} C_{T_{\mu\nu}T_{\rho\sigma}}[k]$$

$$C_{T_{\mu\nu}T_{\rho\sigma}} [k] = {\hbar^2 \over 40\pi}
\theta(k_0) \theta(k^2) (k^2)^2
\pi_{\mu\nu\rho\sigma}$$
\beqa
\label{pi}
\pi_{\mu\nu\rho\sigma} &=& {1\over2}(\pi_{\mu\rho}\pi_{\nu\sigma} +
\pi_{\mu\sigma}\pi_{\nu\rho}) -
{1\over3}\pi_{\mu\nu}\pi_{\rho\sigma}\nonumber\\
\pi_{\mu\nu} &=& \eta_{\mu\nu} - {k_\mu k_\nu \over k^2}\nonumber
\eeqa
Stress tensor correlations are gauge independent and
 are in fact completely determined, up to a numerical factor,
by general symmetries that are satisfied by correlation functions in
vacuum. Translation and Lorentz invariances imply that correlations in
momentum domain are tensors built from $k_\mu$ and $\eta_{\mu\nu}$ only.
Correlations of
stress tensors $T_{\mu\nu}$ and $T_{\rho\sigma}$ decompose on
tensors which are symmetric in indices ($\mu,\nu$), ($\rho,\sigma$) and
exchange of these pairs, and which are transverse because of
energy-momentum conservation ($\partial^\mu T_{\mu\nu} = 0$). As Maxwell
stress tensor is moreover traceless, this leaves only one such tensor
$\pi_{\mu\nu\rho\sigma}$. In momentum domain, stress tensor correlations
are obtained as convolutions of field correlation functions in vacuum, so
that they only contain time-like momenta (given by adding two light-like
momenta of positive frequencies), and thus a
factor $\theta(k_0)\theta(k^2)$. Quite generally, as discussed at the end of
this section, such factor can also be seen
as a consequence of fluctuation-dissipation relations
characteristic of vacuum, and of Lorentz invariance.
By dimensionality, these correlations are proportional to $(k^2)^2$.
Explicit computation provides the remaining numerical factor.

When seen as a boundary condition for the field, a perfect mirror determines
a relation between outcoming and incoming fields, which allows one to
derive the fluctuations of radiation pressure exerted on the mirror
from those of incoming fields. Explicit computations have been performed
for mirrors of different shapes
\cite{Barton,Eberlein}.

\bigskip
We shall briefly discuss the model of partially transmitting mirror
in two-dimensional space-time. The force exerted on the mirror is
given by the difference of field energy densities between the two sides
of the mirror (\ref{f}).
For a single mirror and vacuum input fields, its mean value
vanishes. However, the radiation pressures exerted on both sides of the mirror
have independent fluctuations, and the resulting force still fluctuates.
Force fluctuations on a mirror at rest
are stationary and determined by the mirror's scattering matrix (\ref{S})
and input field correlations (\ref{vac}) \cite{JR2}:
$$<F(t)F(t')> = C_{FF}(t-t')$$
\beqn
\label{ff}
C_{FF}[\omega] = 2 \hbar^2 \theta(\omega)
\int_0^\omega {d\omega' \over 2\pi}
\omega'(\omega-\omega')
\rm{Re}\lbrace 1 - s[\omega']s[\omega-\omega'] +
r[\omega']r[\omega-\omega']\rbrace
\eeqn
Force correlations are positive and always finite. They vanish as
$\omega^3$ around zero frequency (energy-momentum is conserved
in vacuum). In particular
for a perfect mirror, they are directly related to correlations of momentum
 densities of incoming free fields (see (\ref{bc})):
\beqn
\label{df}
C_{FF}[\omega] =  {\hbar^2 \over 3\pi} \theta(\omega) \omega^3
\eeqn
Correlation spectra in vacuum contain positive frequencies only.
Force correlations can similarly be derived for thermal fields. In thermal
equilibrium, a fluctuation-dissipation relation
relates commutators and anticommutators \cite{Kubo}:
\beqa
2\hbar \xi_{FF}(t) &=& <[F(t), F(0)]>\nonumber\\
2\hbar \sigma_{FF}(t) &=& <\lbrace F(t), F(0) \rbrace>\nonumber
\eeqa
\beqn
\label{fd1}
2\hbar \xi_{FF}[\omega]
= (1 - e^{-{\hbar \omega \over T}}) C_{FF}[\omega]
\eeqn
$$\sigma_{FF}[\omega] = \rm{coth} {\hbar\omega \over 2T} \xi_{FF}[\omega]$$
($T$ is the temperature).
This fluctuation-dissipation relation, which was first studied for
Nyquist noise in electric circuits \cite{CW}, allows
to determine fluctuations in a state of thermal equilibrium
from the commutator only.
As discussed in next section, the commutator
also identifies with the dissipative part of the linear
response of the system to an external perturbation \cite{Kubo}.
At the limit of zero temperature,
that is in vacuum, the fluctuation-dissipation relation leads to
a fluctuation spectrum which is limited to positive frequencies:
\beqn
\label{fd3}
C_{FF}[\omega] = 2\hbar \theta[\omega] \xi_{FF}[\omega]
\eeqn
The factor $\theta[\omega]$ explicitly shows that correlations (\ref{ff})
are not symmetric under exchange of time arguments and thus
exhibits the non-commutative character (i.e.
the quantum nature) of fluctuations in vacuum.

\bigskip
\begin{flushleft}
{\bf 4 Radiation reaction force}
\end{flushleft}

As first discussed by Einstein,
a scatterer immersed in a fluctuating field undergoes a Brownian motion
\cite{E2}. The fluctuating force exerted by the field leads to a
diffusion process for the scatterer's momentum $P$,
which spreads in time according to:
$$<\Delta P^2> \sim 2 D \Delta t$$
The momentum distribution being constant at thermal equilibrium,
the effect of the fluctuating
force must be exactly compensated by the effect of a further cumulative
force which appears when the scatterer moves (with velocity $\delta {\dot q}$):
$$<\delta F > \sim - \gamma \delta {\dot q}$$
This implies a relation
between the momentum diffusion coefficient $D$ which characterises force
fluctuations and the friction coefficient $\gamma$ which characterises the
mean dissipative force:
\beqn
\label{fd0}
D = \gamma T
\eeqn
($T$ is the temperature). In the general quantum case, when small
displacements of the scatterer are considered,
a relation still holds between force fluctuations exerted on
the scatterer at rest and the mean dissipative force exerted on
the moving scatterer, i.e. the imaginary part of the
force susceptibility $\chi_{FF}$:
\beqn
\label{fs}
<\delta F[\omega]> = \chi_{FF}[\omega] \dq[\omega]
\eeqn
Indeed, according to linear response theory \cite{Kubo},
the susceptibility of a quantity under a
perturbation is related to the correlations, in the unperturbed state,
of this quantity with the generator of the perturbation.  More precisely,
fluctuation-dissipation relations identify the imaginary (or
dissipative)
part of the susceptibility of a quantity with its commutator with the
generator of the perturbation.
In the case of displacements, the generator is given by the force
itself:
\beqn
\label{fd2}
  \rm{Im} \chi_{FF}[\omega]
= \xi_{FF}[\omega]
\eeqn
In a thermal state, according to fluctuation-dissipation relations (\ref{fd1})
and (\ref{fd2}), force fluctuations are connected with the mean motional force:
$$2 \rm{Im} \chi_{FF}[\omega] =
{1\over \hbar}(1 - e^{-{\hbar \omega \over T}}) C_{FF}[\omega]$$
At the limit of high temperature, Einstein's relation
is recovered (see \ref{fd0}):
$$2 \rm{Im} \chi_{FF}[\omega] = {\omega \over T} C_{FF}[\omega]$$
At the limit of zero temperature, force fluctuations subsist
on a scatterer at rest and imply that, when moving in vacuum, a scatterer
is submitted to a dissipative force. The motional susceptibility of the force
is then determined by force correlations in vacuum (\ref{ff}) and
fluctuation-dissipation relation (\ref{fd2}). For a perfect
mirror, at first order in the mirror's
displacement, the mean dissipative force is proportional to
the third time derivative (see (\ref{df})):
\beqn
\label{rrf}
<\delta F(t)> =  { \hbar \over 6 \pi } \dddq(t)
\eeqn

More recently, effects of moving boundaries on quantum fields have been
discussed to study the influence of classical
constraints on quantum fields, and in particular as an analogy for quantum
fields in a classical curved space \cite{deWitt}. A perfect mirror moving
in vacuum has been shown to radiate and hence to undergo a radiation reaction
force. In two-dimensional space-time, and for scalar fields in vacuum,
the radiation reaction force is proportional to the two-dimensional version
of Abraham vector \cite{FD}, and identifies with (\ref{rrf}) in linearised and
non relativistic limits. The radiation reaction force for partially
transmitting mirrors can also be obtained following this approach. A moving
mirror induces a motional modification of the field scattering matrix
(\ref{S}).
Using a coordinate transformation to the mirror's proper frame, the
modification
of the scattering matrix is easily obtained up to first order in the mirror's
displacement \cite{JR2}:
\beqn
\label{msm}
\left(\matrix{\delta\varphi_{out}[\omega]\cr
\delta\psi_{out}[\omega]\cr}\right) =
\int {d\omega'\over2\pi} \delta S[\omega,\omega']
\left(\matrix{\varphi_{in}[\omega']\cr
\psi_{in}[\omega']\cr}\right)
\eeqn
For general motions of the mirror, vacuum incoming fields are transformed
into outcoming fields whose correlations do not correspond to vacuum. Then,
energy-momentum is radiated by the moving mirror. As the mean stress
tensor of outcoming fields differs from that of incoming fields,
the radiation pressure exerted by scattered fields does not vanish and the
mirror is submitted to a radiation reaction force (\ref{fs}). The radiation
pressure exerted by scattered fields (see (\ref{f})) can be obtained using the
modified scattering matrix (\ref{msm}) \cite{JR2}:
\beqn
\label{xff}
\chi_{FF}[\omega] = i\hbar \int_0^\omega {d\omega'\over2\pi} \omega'(\omega -
\omega') \lbrace 1 - s[\omega']s[\omega-\omega'] +
r[\omega']r[\omega-\omega']\rbrace
\eeqn
As expected, this expression and (\ref{ff}) satisfy fluctuation-dissipation
relation (\ref{fd2}). As a consequence of Lorentz invariance of vacuum, the
radiation reaction force vanishes for uniform motion; expression (\ref{xff})
leads to (\ref{rrf}) for a perfect mirror and also vanishes for
uniformly accelerated motion:
$$\chi_{FF}'[0] = \chi_{FF}^{''}[0] = 0$$

The radiation reaction force felt by a plane perfect mirror in four-dimensional
space-time has been obtained for motions in scalar field \cite{FV} and in
electromagnetic \cite{Maia Neto} vacua.
Direct comparison between radiation pressure fluctuations and the dissipative
force shows that fluctuation-dissipation relations
are satisfied in these cases.

\bigskip
\begin{flushleft}
{\bf 5 Stability of motion and position fluctuations in vacuum}
\end{flushleft}

It is well-known from classical electron theory \cite{Rohrlich}, that a
radiation reaction force proportional to Abraham-Lorentz vector (\ref{rrf})
leads to motions which are either unstable
(the so-called "runaway solutions") or violate causality.
Perfect mirrors, and in
particular mirrors treated as field boundaries, have motions in vacuum
which are affected by the same stability and causality problems.
However, as we briefly show in the following, partially transmitting
mirrors can avoid these difficulties \cite{JR3}.

When a scatterer is submitted to an applied external force $F(t)$, its motion
is determined by an equation which also takes into account the force
induced by motion (\ref{fs}). It will be sufficient for our purpose
to consider small displacements only, so that the equation of motion
reads (we generally consider a mirror of mass $m_0$,
bound with a proper frequency
$\omega_0$; a free mirror is recovered for $\omega_0 = 0$):
\beqn
\label{leq}
m_0 ({\ddot q}(t) + \omega_0^2 q(t)) = F(t) + \int_{-\infty}^\infty dt'
 \chi_{FF}(t-t') q(t')
\eeqn
The motional response is best characterised in the frequency domain,
by a mechanical impedance $Z[\omega]$ which relates the resulting velocity
to the applied force:
\beqn
\label{rep}
-i\omega Z[\omega] q[\omega] = F[\omega] \qquad \qquad
Z[\omega] = -im_0\omega + i{m_0 \omega_0^2\over \omega} +
{\chi_{FF}[\omega] \over i \omega}
\eeqn
{}From analytic properties of the scattering matrix (\ref{S}), it is easily
derived that $\chi_{FF}$ is also analytic in the upper half plane
($\rm{Im}(\omega) > 0$), showing that the induced force is a causal function
of the scatterer's motion.
The mechanical impedance of a perfect mirror is obtained from (\ref{rrf}):
\beqn
Z[\omega] = -im_0\omega + i{m_0 \omega_0^2\over \omega} +
 { \hbar \over 6 \pi} \omega^2 \nonumber
\eeqn
As easily seen, the mechanical admittance ($Y[\omega] = Z[\omega]^{-1}$) is no
longer analytic but has a pole in the upper half plane. Such a pole
corresponds to unstable motions, i.e. "runaway solutions" in the free
mirror case. If supplementary boundary conditions are imposed
(at large time for instance) to exclude such solutions, then the resulting
motions violate the causal dependence on the applied force.

A partially transmitting mirror is characterised by a force
susceptibility of the general form:
$$\chi_{FF}[\omega] = i{ \hbar \over 6 \pi} \omega^3 \Gamma[\omega]$$
High frequency transparency and analyticity lead to a better behaviour
of the susceptibility \cite{JR3}:
$$\Gamma[\omega] \sim i{\omega_c \over \omega}
\qquad  \rm{for} \qquad \omega \rightarrow \infty$$
where $\omega_c$ defines a high frequency cut-off.
Characteristic relations in vacuum derived from
(\ref{fd3}) and (\ref{fd2}) furthermore imply:
\beqn
\rm{Im}({\chi_{FF}[\omega]\over \omega}) \ge 0 \qquad
\rm{for} \qquad \rm{Im}(\omega) \ge 0
\nonumber
\eeqn
This property means that the force susceptibility
is related to a positive function \cite{Meixner}. The mechanical
impedance defined by (\ref{rep}) corresponds to a positive function
as long as its pole at infinity satisfies:
\beqn
\label{res}
m_\infty = m_0 -
 { \hbar \omega_c\over 6 \pi } \ge 0
\eeqn
The inverse
of a positive function is still a positive function, and furthermore
causality follows from positivity \cite{Meixner}.
When inequality (\ref{res}) is satisfied, the mechanical
admittance is also a causal function,
and no "runaway solutions" can appear (see (\ref{rep})).
$m_0$ and $m_\infty$ can be seen as describing a low and a high frequency mass
for the mirror, their difference being a mass correction induced by
the field interacting with the mirror. Stability in vacuum is insured as
long as the free mass of the mirror $m_\infty$ is positive, or else as long
as the quasistatic mass of the mirror $m_0$ is larger than the induced mass.
Realistic mirrors certainly satisfy this condition,
contrarily to perfect mirrors. This discussion also
shows the incompatibility between stability in vacuum
and mass renormalisation, since $m_\infty$ is infinitely negative in that
case (see \cite{Dekker} for a similar discussion for electrons).

\bigskip
While motion of the mirror modifies the scattered field, the radiation
pressure exerted by the field also perturbs the mirror's motion. Equations
of the coupled system, mirror and vacuum field radiation pressure, when
treated linearly, can be solved to obtain fluctuations for the interacting
system in terms of input fluctuations only \cite{JR5}. In particular,
equation (\ref{leq}) provides the mirror's position fluctuations
in terms of input force fluctuations.
As a consequence of consistency relations satisfied by linear response
formalism, position fluctuations of the mirror coupled to vacuum fields
also satisfy fluctuation-dissipation relations,
which relate the position commutator to the dissipative
part of the mechanical admittance (the mirror's response to an applied force):
\beqa
\label{y}
Y[\omega] &=& {-i \omega \over
m_0(\omega_0^2 - \omega^2) - \chi_{FF}[\omega]}\nonumber\\
\xi_{qq}[\omega]  &=& \rm{Re} Y [\omega] / \omega
\eeqa
Fluctuations of the coupled system also satisfy the relations characteristic
of vacuum:
$$C_{qq}[\omega] = 2 \hbar \theta(\omega) \xi_{qq}[\omega]$$
The mechanical admittance completely determines position fluctuations in
vacuum.
As shown by expression (\ref{y}), position fluctuations consist of two main
parts. One part corresponds to a resonance peak at the
oscillator's proper frequency $\omega_0$.
These fluctuations subsist at the limit of
decoupling between mirror
and field and can be seen as proper position fluctuations of the mirror; they
identify with the fluctuations associated with Schr\"odinger equation for
a free oscillator.
The other part is a background noise spreading over all frequencies.
These are position fluctuations induced by the
fluctuating radiation pressure of vacuum fields and are dominant outside
resonance peaks; they describe ultimate position fluctuations and correspond to
a quantum limit on the sensitivity of an optimal position measurement
performed at such frequencies \cite{Braginsky,JR5}.

\bigskip
\begin{flushleft}
{\bf 6 Conclusion}
\end{flushleft}

Quantum field fluctuations in vacuum exert a fluctuating radiation pressure
on scatterers. In agreement with fluctuation-dissipation relations, a
scatterer moving in vacuum experiences an additional force depending on
its motion. Like in classical electron theory, it generally modifies
the scatterer's motional response to an applied force.
For realistic
mirrors, which are transparent to high frequencies,
 motions can be shown to remain
stable and causal in vacuum, a property which is violated by mass
renormalisation.

The quantum Brownian motion induced on the mirror's position by its coupling
to vacuum field radiation pressure can be described consistently within
linear response formalism.
Position fluctuations are determined from the mechanical susceptibility
of the mirror through fluctuation-dissipation relations. Vacuum field
fluctuations then lead to ultimate quantum fluctuations for positions in
space-time.


\bigskip
\begin{thebibliography}{99}
\bibitem{Lorentz}
H.A. Lorentz, "The Theory of Electrons" (Leipzig, 1915) [reprinted by
Dover, New York, 1952]
\bibitem{E1}
A. Einstein, Ann. Physik {\bf 18} (1905) 639, {\bf } (1906) 627;
Jahrb. Radioakt. Elektron. {\bf 4} (1907) 411;{\bf 5} (1908) 98
[translated in English and commented by H.M. Schwartz,
Am. J. Phys. {\bf 45} (1977) 512, 811, 899].
\bibitem{QED}
H. Groch and E. Kazes, in "Foundations of Radiation Theory and QED",\\
A.O. Barut ed. (Plenum Press, New York, 1980).
\bibitem{BR}
N. Bohr and L. Rosenfeld, Phys. Rev. {\bf 78} (1950) 794.
\bibitem{Braginsky}
V.B. Braginsky and F.Ya. Khalili, "Quantum Measurement" (Cambridge, 1992);\\
M.T. Jaekel and S. Reynaud, Europhys. Lett. {\bf 13} (1990) 301;
Phys. Lett. {\bf A 185}  (1994) 143.
\bibitem{Casimir}
H.B.G. Casimir, Proc. K. Ned. Akad. Wet. {\bf 51} (1948) 793;\\
G. Plunien, B. M\"uller and W. Greiner, Phys. Rep. {\bf 134}
(1986) 87.
\bibitem{E2}
A. Einstein, Ann. Physik {\bf 17} (1905) 549;
Phys. Z. {\bf 10} (1909) 185,
{\bf 18} (1917) 121.
\bibitem{Wesson}
P.S. Wesson, Astrophys. J. {\bf 378} (1991) 466.
\bibitem{BM}
L.S. Brown and G.J. Maclay, Phys. Rev. {\bf 184} (1969) 1272.
\bibitem{JR1}
M.T. Jaekel and S. Reynaud,  J. Phys. I France {\bf 1} (1991)
1395.
\bibitem{Barton}
G. Barton, J. Phys. A: Math.Gen. {\bf 24} (1991) 991, 5533;
in "Cavity Quantum
Electrodynamics" (Supplement: Advances in Atomic, Molecular and
Optical Physics) P. Berman ed. (Academic Press, 1994).
\bibitem{Eberlein}
C. Eberlein, J. Phys. A: Math.Gen. {\bf 25} (1992) 3015, 3039.
\bibitem{JR2}
M.T. Jaekel and S. Reynaud, Quantum Optics {\bf 4} (1992)
39.
\bibitem{Kubo}
R. Kubo, Rep. Prog. Phys. {\bf 29} (1966) 255.\\
L.D. Landau and E.M. Lifschitz, Cours de  Physique Th\'eorique,
Physique Statistique, premi\`ere partie
(Mir, Moscou, 1984) ch. 12.
\bibitem{CW}
H.B. Callen and T.A. Welton, Phys. Rev. {\bf 83} (1951) 34.
\bibitem{deWitt}
B.S. de Witt, Phys. Rep. {\bf 19} (1975) 295.\\
N.D. Birell and P.C.W. Davies, "Quantum Fields in Curved Space"
(Cambridge, 1982);\\
S.A. Fulling, "Aspects of Quantum Field Theory in Curved Spacetime"
(Cambridge,1989).
\bibitem{FD}
S.A. Fulling and P.C.W, Davies, Proc. R. Soc. {\bf A 348} (1976) 393.
\bibitem{FV}
L.H. Ford and A. Vilenkin, Phys. Rev. {\bf D 25} (1982) 2569.
\bibitem{Maia Neto}
P.A. Maia Neto, J. Phys. A Math. Gen. {\bf 27} (1994) 2167.
\bibitem{Rohrlich}
F. Rohrlich, "Classical Charged Particles"
(Addison-Wesley, Reading, MA, 1965).
\bibitem{JR3}
M.T. Jaekel and S. Reynaud, Phys. Lett. {\bf A 167} (1992) 227.
\bibitem{Meixner}
J. Meixner, in "Statistical Mechanics of Equilibrium and Non-Equilibrium"
J. Meixner ed. (North-Holland, Amsterdam, 1965).
\bibitem{Dekker}
H. Dekker, Phys. Lett. {\bf 107 A} (1985) 255.\\
        Physica {\bf 133 A} (1985) 1.
\bibitem{JR5}
M.T. Jaekel and S. Reynaud, J.Phys.I France  {\bf 3} (1993) 1.

\end{thebibliography}
\end{document}